\documentclass[aps,prd,showkeys,nofootinbib,superscriptaddress,preprint]{revtex4-1}
 \usepackage{mathtools,dsfont,tensor,epsf}
 \usepackage{graphicx}

\usepackage{adjustbox}

\usepackage[top=2cm, bottom=3cm, left=3cm, right=2cm]{geometry}

\usepackage{hyperref}
\usepackage{color}
\definecolor{darkgreen}{rgb}{0,0.35,0}

\newcommand{\mk}{ |\vec{k}|}
\usepackage{amsfonts}
\usepackage{amsmath}
\usepackage{subfig}
\usepackage{mathtools,slashed,tensor}

\newcommand{\ucharles}{Faculty of Mathematics and Physics, Charles University, V Hole\v{s}ovi\v{c}k\'ach 2, 18000 Prague 8, Czech Republic}
\newcommand{\kulak}{Department of Physics, KU Leuven Campus Kortrijk--Kulak, Etienne Sabbelaan 53 bus 7657, 8500 Kortrijk, Belgium}

\newcommand{\infn}{ INFN, Sezione di Napoli, Complesso Universitario di Monte S.~Angelo, Via Cintia Edificio 6, 80126 Naples, Italia}
\newcommand{\napoli}{Dipartimento di Matematica e Applicazioni "R. Caccioppoli", Universit\'{a} di Napoli Federico II, Complesso Universitario di Monte S.~Angelo,  Via Cintia Edificio 6,
80126 Naples, Italia }
\newcommand{\uach}{Instituto de Ciencias F\'isicas y Matem\'aticas, Universidad Austral de Chile, Casilla 567, 5090000 Valdivia, Chile.}
\newcommand{\ughent}{Ghent University, Department of Physics and Astronomy, Krijgslaan 281-S9, 9000 Ghent, Belgium}

\begin{document}

\title{The Casimir energy in terms of boundary quantum field theory: the QED case
}

\author{David Dudal}
\email{david.dudal@kuleuven.be}
\affiliation{\kulak}
\affiliation{\ughent}
\author {Pablo Pais}\thanks{On leave from Institute of Physics of the ASCR, ELI Beamlines Project, Na Slovance 2, 18221 Prague, Czech Republic.\\}
\email{pais@ipnp.troja.mff.cuni.cz}
\affiliation{\uach}
\affiliation{\ucharles}
\author{Luigi Rosa}
\email{rosa@na.infn.it}
\affiliation{\napoli}
\affiliation{\infn}


\begin{abstract}
We revisit the path integral computation of the Casimir energy between two infinite parallel plates placed in a QED vacuum. We implement perfectly magnetic conductor boundary conditions (as a prelude to the dual superconductor picture of the QCD vacuum) via constraint fields and show how an effective gauge theory can be constructed for the constraint boundary fields, from which the Casimir energy can be simply computed, in perfect agreement with the usual more involved approaches. Gauge invariance is natural in this framework, as well as the generalization of the result to $d$ dimensions. We also pay attention to the case where the outside of the plates is not the vacuum, but a perfect magnetic (super)conductor, disallowing any dynamics outside the plates. We find perfect agreement between both setups.
\end{abstract}

\maketitle

\section{Motivation}

The Casimir energy and its related force per unit area \cite{Casimir:1948dh} are  amongst the most spectacular effects of the quantum vacuum not being ``empty'': thanks to the non-trivial virtual particles swarming between two (electromagnetically uncharged) parallel (flat) plates, these can attract each other. This can also be traced back to the boundary conditions the quantum electromagnetic field modes are subject to, see e.g.~\cite{Plunien:1986ca,Bordag:2001qi,Milton:2001yy}. Experimental evidence for this (tiny) effect in the original (plane-plane) configuration
can be found in e.g.~\cite{Bressi:2002fr}, while an extensive report of the various experiments made  over the years can be found in \cite{Bordag:2001qi}, and \cite{Lambrecht:1999vd}. Since the seminal work \cite{Casimir:1948dh}, the Casimir effect has been studied in a variety of field theories and for variable geometries. We refer to the aforementioned reviews \cite{Plunien:1986ca,Bordag:2001qi,Milton:2001yy} for more details.

During recent years, one witnessed an increased interest in the relevance of so-called edge (boundary) modes in high energy physics\footnote{See \cite{Gorbar:2001qt,Teber:2018jdh,Dudal:2018pta,Donnelly:2015hxa,Gomes:2018shn,Blommaert:2018oue,Herzog:2017xha,DiPietro:2019hqe,Kurkov:2020jet} for a deliberately short illustrative selection.}, an issue also well known from the condensed matter community \cite{Qi:2011zya}.

Given the intimate connection between the Casimir effect and boundary conditions, one cannot help but wonder whether the Casimir effect cannot be understood from a type of ``boundary dynamics'', and the answer is indeed affirmative as we will discuss in this note, inspired by the work \cite{Bordag:1983zk}.

We will first give a short survey about boundary conditions for (Abelian) gauge theories in the gauge fixed setting, with special attention being paid to the issue of gauge invariance. Indeed, this is by far the most delicate issue when dealing with boundary conditions: how to impose these without jeopardizing the local gauge invariance of the theory. After that, we will derive the effective boundary action, followed by a different route where the boundary conditions are imposed before the gauge fixing, leading to the very same effective action and ensuing Casimir energy/force. The latter methodology will allow for a simpler generalization to non-Abelian gauge theories at a later stage.  Before the conclusions, we redo the latter exercise in case the plate geometry is enclosed in a perfect (magnetic) conductor, i.e.~we solve for the Casimir force per unit area in a periodic setting. We shall see that the Faddeev-Popov (FP) ghosts play an important r\^ole in the latter periodic approach, but not in the former non-periodic computation. We do however find the same Casimir force per unit area. To our knowledge, this is a priori a non-trivial result not really discussed in present literature. We will end with the generalization to $d$ dimensions and outlook to further research. We include a series of technical Appendices.

\section{Maxwell action and boundary conditions}

Let us take the classical Euclidean action on a $4-$dimensional manifold ${\cal M}$ with boundary $\Sigma$
\begin{equation}\label{action_Maxwell}
\int_{\cal{ M}}{\cal{ L}}_{\text{E}}=\int_{\cal{ M}} \frac{1}{4}\left(F_{\mu\nu}F_{\mu\nu}\right) \;,
\end{equation}
where, as usual, $F_{\mu\nu}=\partial_{\mu}A_{\nu}-\partial_{\nu}A_{\mu}$ is the Maxwell field strength tensor. As usual in quantum field theory, in order to obtain the field equations, one takes the variation, in the functional sense, of the action whilst terms obtained at the (infinitely far) boundary are dropped by requiring that the fields decay to zero fast enough. More care must be taken, when the manifold has boundaries, to ensure that the action is an extremum when the bulk field equations hold\footnote{Even if the boundary is far away or at infinity, general relativity teach us the lesson that boundary terms cannot be neglected. In fact, in such situations one need to take explicit care of them to define properly the mass, angular momentum and possible net charge of a black hole.}. Therefore, let us variate the action \eqref{action_Maxwell}, but this time we keep track of the boundary terms coming from partial integration. So,
\begin{equation*}
\delta S_{\text{E}} = \int_{\mathcal{M}} \delta A_{\mu}\partial_{\nu}F_{\mu\nu}  - \int_{\Sigma} n_{\nu}\delta A_{\mu}F_{\mu\nu} \;,
\end{equation*}
where $\Sigma$ is the boundary of $\mathcal{M}$, and $n_{\mu}$ is a unit 4-vector normal to $\Sigma$.

By looking at the extremum of $S_{\text{E}}[A]$ with respect to an arbitrary variation of $A_\nu$, i.e.~$\delta S_{\text{E}}=0$, this leads to the field equations and boundary conditions, respectively,
\begin{eqnarray}
 \partial_\nu F_{\mu\nu} &=&0 \;,\quad \label{Maxwell_vacuum} n_{\mu}F_{\mu\nu} \, \Big{|}_{\Sigma} =0   \;.
\end{eqnarray}
The second equation is the dual superconductor boundary condition (DSBC) \cite{Bordag1998}, also known as ``perfect magnetic boundary conditions'' (PMC). It is important to stress at this point that the DSBC is obtained naturally as the requirement that the action \eqref{action_Maxwell} leads to a genuine extremum in the presence of boundaries. Said otherwise, the boundary conditions follow from the action principle and are not put ```by hand'' on top of the (quantum) equations of motion. Let us mention here that boundary conditions are also indispensable to ensure Hermiticity.

As we know, the Maxwell action has a gauge redundancy when $A_{\mu} \to A'_{\mu}=A_{\mu}+\partial_{\mu}\alpha$, for an arbitrary space-time function $\alpha(x)$. We must add a gauge fixing term to the action
\begin{equation*}
S_{\text{E}}=\int d^4x\left(\frac{1}{4}F_{\mu\nu}F_{\mu\nu} + \frac{1}{2\xi}\partial_{\mu}A_{\mu}\partial_{\nu}A_{\nu}\right) \;,
\end{equation*}
where $\xi$ is a gauge fixing parameter that could be any finite value. Physical observables should not depend on this $\xi$. From the variation of the action with respect to the field $A_{\mu}$,
\begin{equation}
\int_{\mathcal{M}} \delta A_{\mu}(\partial_{\nu}F_{\mu\nu}-\frac{1}{\xi}\partial_{\mu}\partial_{\nu}A_{\nu})  - \int_{\Sigma} n_{\nu}\delta A_{\mu}(F_{\mu\nu}-\frac{\delta_{\mu\nu}}{\xi}\partial_{\rho}A_{\rho}) \;. \label{variation_action_xi}
\end{equation}
we now obtain as the field equations and boundary conditions, respectively,
\begin{eqnarray}
 \partial_\nu F_{\mu\nu} - \frac{1}{\xi}\partial_{\mu}\partial_{\nu}A_{\nu}  &=&0 \,,\quad \label{Maxwell_vacuum_xi}
n_{\nu}F_{\mu\nu} - \frac{1}{\xi} {n_{\mu}}\partial_{\rho}A_{\rho} \, \Big{|}_{\Sigma} =0   \;.
\end{eqnarray}
Because of the antisymmetry of $F_{\mu\nu}$, we derive the first equation w.r.t.~$x_\mu$ and multiply the second by $n_{\mu}$, to get
\begin{eqnarray*}
&&\partial^2\partial_{\nu}A_{\nu}  =0 \;,  \partial_{\nu}A_{\nu} \, \Big{|}_{\Sigma} =0  \Rightarrow \partial_{\nu}A_{\nu}  =0 \;,  \partial_{\nu}A_{\nu} \, \Big{|}_{\Sigma} =0   \;,\end{eqnarray*}
as $1/\xi$ is not zero. By adding the gauge fixing term, we ensure on-shell that $\partial_{\nu}A_{\nu}=0$, including at the boundary $\Sigma$ of course.

Although the Faddeev-Popov (FP) ghost and anti-ghost fields, $c$ and $c^{\dagger}$, are usually said to be unnecessary for the Maxwell theory's quantization, it is known that they can be important when boundaries are present, even in Abelian theories \cite{Ambjorn1982}. Therefore, we will introduce these Grassmann scalar fields to get the following action
\begin{equation}\label{Lagrangian}
S_{\text{E}}=\int d^4x \left(\frac{1}{4}F_{\mu\nu}F_{\mu\nu}+\frac{1}{2\xi}(\partial_{\mu}A_{\mu})^{2}-\partial_{\mu}c^{\dagger}\partial_{\mu}c\right) \;.
\end{equation}
If the boundary conditions \eqref{Maxwell_vacuum_xi} are satisfied, \eqref{Lagrangian} is invariant under both BRST,
\begin{eqnarray}\label{BRST}
s A_{\mu} &=& -\partial_{\mu} c\,,\quad s c^{\dagger} = \frac{1}{\xi}\partial_{\mu}A_{\mu}\,,\quad s c = 0 \;, \nonumber
\end{eqnarray}
and anti-BRST \cite{Ojima1980}
\begin{eqnarray}\label{anti_BRST}
s^{\dagger} A_{\mu} &=& -\partial_{\mu} c^{\dagger}\,,\quad s^{\dagger} c = - \frac{1}{\xi}\partial_{\mu}A_{\mu}\,,
\quad s^{\dagger} c^{\dagger} = 0 \;, \nonumber
\end{eqnarray}
transformations\footnote{The boundary condition $\partial_{\mu}A_{\mu}|_{\Sigma}=0$ is necessary because these (on-shell) BRST and anti-BRST transformations are not nilpotent (e.g.~$s^2\neq0$). To be nilpotent, we would need to introduce an auxiliary Nakanishi-Lautrup scalar field $h$, which complicates the method we pursuit here. We will however do this later.}. As the field equations for the ghost and anti-ghost fields are, respectively,
$\partial^{2} c = 0$,  $\partial^{2} c^{\dagger} = 0$, the boundary conditions \eqref{Maxwell_vacuum_xi} are BRST and anti-BRST invariant without imposing further boundary conditions on the ghost and anti-ghost fields \cite{Ambjorn1982}.

\section{Boundary conditions for parallel plates}

\begin{figure}
\begin{center}
\includegraphics[width=0.25\textwidth,angle=0]{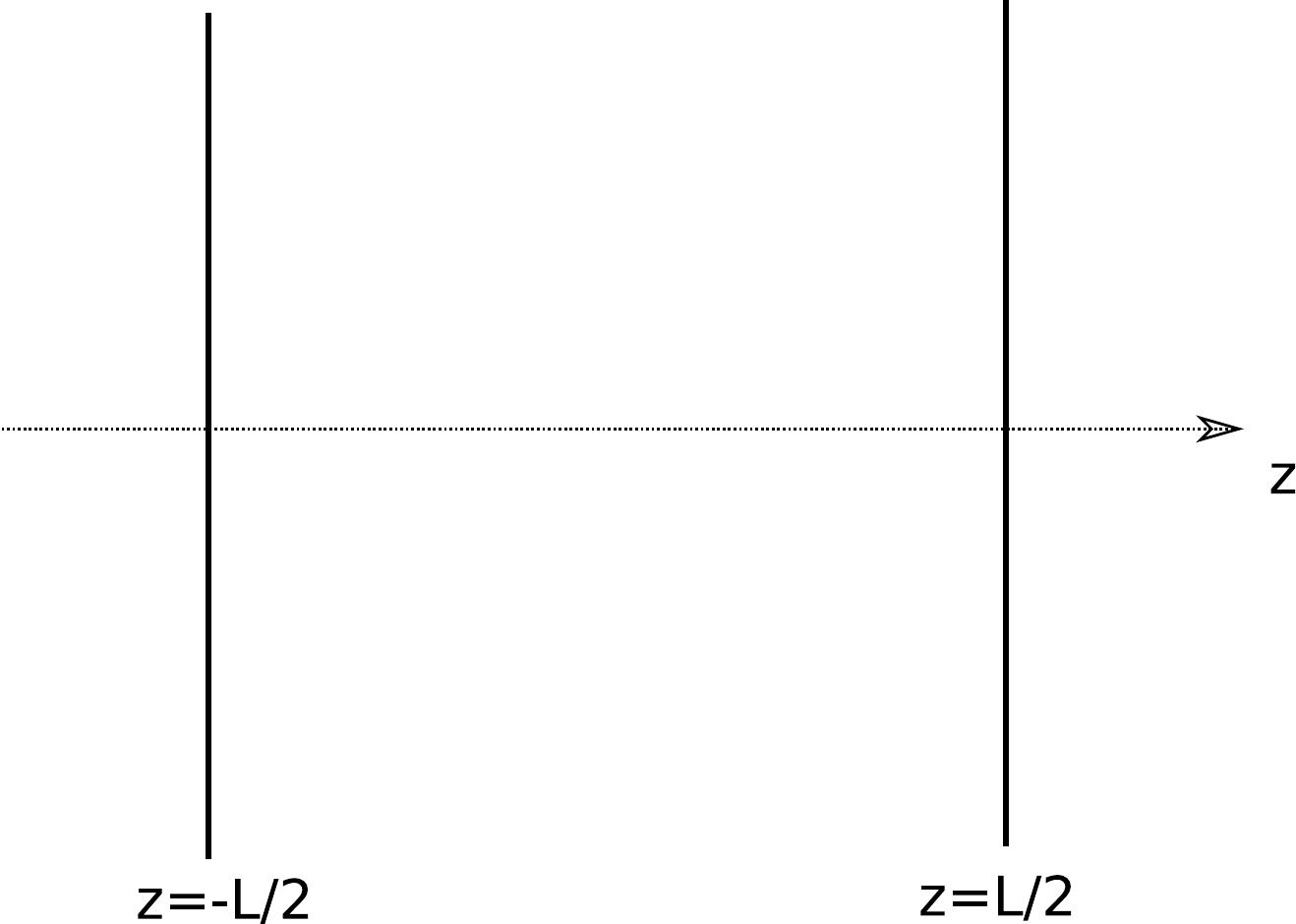}
\end{center}
\caption{{\protect\small Representation of two infinite parallel plates localized at $z=\pm L/2$. The dual superconductor boundary conditions \eqref{boundary_condition} imply that the perpendicular electric field component $E_{z}$ and parallel magnetic field components $B_{x}\;,B_{y}$ are zero on each plate at $z=\pm L/2$. }}%
\label{parallel_planes}%
\end{figure}

All of the above discussion can now be applied for any space-time manifold $\cal M$ with a general boundary $\Sigma$. Here, we consider a $(3+1)$-dimensional space-time where two parallel infinite planes localized at $z=\pm L/2$ acts as the boundary (see FIG.~\ref{parallel_planes}). Taking the DSBC plus the gauge fixing \eqref{Maxwell_vacuum_xi}, we arrive at
\begin{eqnarray}\label{boundary_condition}
n_{\mu}F_{\mu\nu}\big{|}_{z=\pm L/2} &=& 0 \;,\quad \partial_{\nu}A_{\nu} \, \Big{|}_{z=\pm L/2} = 0 \;.
\end{eqnarray}
Here, $n_{\mu}=(0,0,0,1)$, being perpendicular to the plates. At this point, it is convenient to split the space-time indices $\mu$ in $z$ and the rest: $x_{\mu}=(\vec{x},z)$, with $\vec{x}=(x_{0},x_{1},x_{2})$ and the Latin index $i\in\{0,1,2\}$.

How could we implement the boundary condition \eqref{boundary_condition} in a suitable effective QED action? As the quantized action is defined in terms of the potential $A_\mu$, one possibility is to deduce from \eqref{boundary_condition} conditions on the potential $A_\mu$ capable to ensure a self-adjoint wave operator and, at the same time, to preserve gauge (BRST) invariance (also on the boundary).
Unfortunately, this is not a trivial task (see, for instance, \cite{Avramidi:1997hy,Witten:2018lgb,Vassilevich:2003xt}).

Another possibility, the one we are adopting here, is to consider boundary conditions as constraints when quantizing the potential $A_\mu$, a setup initiated by \cite{Bordag:1983zk}, see also \cite{Bordag1998,Chernodub:2016owp,Chernodub2018,Karabali:2015epa}. This can be done by adding to the action  auxiliary fields $B_{\mu}(x)$ and $\bar{B}_{\mu}(x)$ that act as Lagrange multipliers. For our case, we take as these auxiliary fields
\begin{eqnarray}
B_{\mu}(x)
          &=&\left(b_{i}(\vec{x})\delta(z-L/2),b_{z}(\vec{x})\delta(z-L/2)\right)\,,\nonumber \\
\bar{B}_{\mu}(x)
          &=&\left(\bar{b}_{i}(\vec{x})\delta(z+L/2),\bar{b}_{z}(\vec{x})\delta(z+L/2)\right) \;.  \label{auxiliary_fields}
\end{eqnarray}
Stressing the fact that the fields $b$ and $\bar{b}$ have only components in Latin indices and depend only on $\vec{x}$, the new action reads
\begin{eqnarray}\label{action}
S_{E}&=&-\frac{1}{2}\int d^{4}x \left[A_{\mu}\left(\delta_{\mu\nu}\partial^{2}-(1-\frac{1}{\xi})\partial_{\mu}\partial_{\nu}\right)A_{\nu}\right.\\
&&\left.\hspace{-0.5cm}-n_{\mu}F_{\mu\nu}B_{\nu}-\frac{1}{\xi}n_{\mu}B_{\mu}\partial_{\nu}A_{\nu}-n_{\mu}F_{\mu\nu}\bar{B}_{\nu}-\frac{1}{\xi}n_{\mu}\bar{B}_{\mu}\partial_{\nu}A_{\nu}\right]\,. \nonumber
\end{eqnarray}
The variation w.r.t.~$A_{\mu}$ and $B_{\mu}(\bar{B}_{\mu})$ gives us the usual Maxwell equations in vacuum \eqref{Maxwell_vacuum} with Landau gauge condition $\partial_{\mu}A_{\mu}=0$, next to the boundary condition \eqref{boundary_condition}, respectively. Note that the fields $B$ and $\bar{B}$ act as an external source ``living'' only on the plates. This approach has, in our opinion, the big advantage of implementing boundary conditions directly on the fields via $F_{\mu\nu}$. In this way we have to worry neither about the problem of gauge invariance on the boundaries, nor about the ghost and anti-ghost fields at the boundary because, as mentioned above, their field equations guarantee the BRST and anti-BRST invariance at the plates.

\section{Non-local effective action in one dimension less}

We first express the action \eqref{action} in momentum space\footnote{Unlike \cite{Bordag:1983zk} that works in configuration space, our momentum space computation turns out to be much simpler, also to uncloak the gauge invariance of the boundary theory.}, by using the Fourier convention given in the Appendix \ref{appendix_section_1}.
\begin{eqnarray}\label{action_Fourier}
S_{E}&=&-\frac{1}{2} \int\frac{d^{3}k}{(2\pi)^{3}}\int\frac{dk_{z}}{2\pi} \left[\widetilde{A}_{\mu}(k)K^{\xi}_{\mu\nu}\widetilde{A}_{\nu}(-k)\right.\\&&\left.\hspace{-1cm} + \left(ik_{z}\widetilde{A}_{i}(k)-ik_{i}\widetilde{A}_{z}(k)\right) \left(\widetilde{b}_{i}(-k) e^{-ik_{z}\frac{L}{2}} + \widetilde{\bar{b}}_{i}(-k) e^{ik_{z}\frac{L}{2}} \right) \right. \nonumber \\
&& \left.\hspace{-1cm}+\left(\frac{i}{\xi}k_{z}\widetilde{A}_{z}(k)+\frac{i}{\xi}k_{i}\widetilde{A}_{i}(k)\right) \left(\widetilde{b}_{z}(-k) e^{-ik_{z}\frac{L}{2}}+\widetilde{\bar{b}}_{z}(-k) e^{ik_{z}\frac{L}{2}}\right)\right]\nonumber \;,
\end{eqnarray}
where
\begin{equation*}
K^{\xi}_{\mu\nu} = \delta_{\mu\nu} k^{2} - \left(1-\frac{1}{\xi}\right)k_{\mu}k_{\nu} \;.
\end{equation*}
Completing the square, we get
\begin{widetext}
\begin{eqnarray*}
\widetilde{A}_{\mu}(k)(k)K^{\xi}_{\mu\nu}\widetilde{A}_{\nu}(-k)+A_{\mu}(k)v_{\mu}(-k)=
\left(A_{\mu}(k)+\frac{1}{2}v_{\rho}(k)(K^{-1})^{\xi}_{\rho\mu}\right)K^{\xi}_{\mu\nu}\left(A_{\nu}(-k)+\frac{1}{2}(K^{-1})^{\xi}{_{\nu\sigma}}v_{\sigma}(-k)\right) -\frac{1}{4}v_{\mu}(k)(K^{-1})^{\xi}_{\mu\nu}v_{\nu}(-k) \;.
\end{eqnarray*}
\end{widetext}
In our case,
\begin{eqnarray}\label{v_field}
v_{i}(k)&=&-ik_{z}\tilde{b}_{i}(k)e^{ik_{z}L/2}-ik_{z}\tilde{\bar{b}}_{i}e^{-ik_{z}L/2}-\frac{i}{\xi}k_{i}\tilde{b}_{z}(k)e^{ik_{z}L/2}\nonumber\\&&-\frac{i}{\xi}k_{i}\tilde{\bar{b}}_{z}e^{-ik_{z}L/2} \;, \nonumber\\
v_{3}(k)&=&ik_{i}\tilde{b}^{i}e^{ik_{z}L/2}+ik_{i}\tilde{\bar{b}}^{i}e^{-ik_{z}L/2}-\frac{i}{\xi}k_{z}\tilde{b}_{z}(k)e^{ik_{z}L/2}\nonumber\\&&-\frac{i}{\xi}k_{z}\tilde{\bar{b}}_{z}e^{-ik_{z}L/2}  \;, \nonumber\\
(K^{-1})^{\xi}{_{\mu\nu}} &=& \frac{\delta_{\mu\nu}}{k^{2}} + \frac{(\xi-1)}{k^{4}}k_{\mu}k_{\nu} \;.
\end{eqnarray}
Now, by using the fact that the transformation $\widetilde{A}_{\mu}(k) \to \widetilde{A}'_{\mu}(k)=A_{\mu}(k)+\frac{1}{2}v_{\rho}(k)(K^{-1})^{\xi})_{\rho\mu}$ keeps the measure invariant, i.e., $\mathcal{D}A=\mathcal{D}A'$, we can write out the generating functional as
\begin{equation}\label{action_split}
Z=\int\mathcal{D}\widetilde{A'} e^{S[\widetilde{A'}]}\int\mathcal{D}\widetilde{b}\mathcal{D}\widetilde{\bar{b}} e^{S[\widetilde{b},\widetilde{\bar{b}}]}=Z_{A}Z_{b} \;.
\end{equation}

The action \eqref{action} contains both bulk and boundary surface fields, which can be completely separated as written in \eqref{action_split}. The standard approach is to first integrate over the boundary fields, see e.g.~\cite{Bordag:1983zk}, but here we will first integrate over the bulk (gauge) field. The eventual computation of the Casimir energy turns out to be simpler in this fashion relative to \cite{Bordag:1983zk}. A similar ``reversed order'' integration in a slightly different context was also explored in \cite{Bordag:2007zz}.

The next task is to compute explicitly $S[\widetilde{b},\widetilde{\bar{b}}]$, where the dependence on $k_z$ has been integrated out. This can be done because $b$ and $\bar{b}$ do not depend on $k_{z}$. For this purpose, we will make extensive use of the integral formulae in the Appendix \ref{appendix_section_2}. After some computational effort, we arrive at the boundary action we were looking for:
\begin{eqnarray}
 S[\widetilde{b},\widetilde{\bar{b}}] &=&  \frac{1}{2} \int\frac{d^{3}k}{(2\pi)^{3}} \frac{\mk}{8} \left[\widetilde{b}_{i}(k)\left(\delta_{ij}-\frac{k_{i}k_{j}}{\mk^2}\right)\widetilde{\bar{b}}_{j}(-k)e^{-\mk L}\right.\nonumber\\&&\left.\hspace{-1.5cm}+\widetilde{b}_{i}(k)\left(\delta_{ij}-\frac{k_{i}k_{j}}{\mk^2}\right)\widetilde{b}_{j}(-k)+\widetilde{\bar{b}}_{i}(k)\left(\delta_{ij}-\frac{k_{i}k_{j}}{\mk^2}\right)\widetilde{b}_{j}(-k)e^{-\mk L}\right. \nonumber\\
 &+&\left.\widetilde{\bar{b}}_{i}(k)\left(\delta_{ij}-\frac{k_{i}k_{j}}{\mk^2}\right)\widetilde{\bar{b}}_{j}(-k)e^{-\mk L}\right]\;.  \label{action_effective}
\end{eqnarray}
Observe that all the terms containing the gauge parameter $\xi$ neatly canceled out, leading to the promised gauge invariant effective $(2+1)$-dimensional action for the boundary vector fields $b_i$ and $\bar b_i$. For the record, we checked that, conversely, by first integrating over the boundary modes $\widetilde b$ and $\widetilde{\bar b}$ appearing in the action \eqref{action_Fourier}, we recover the momentum space version of the photon propagator in presence of boundaries presented in \cite{Bordag:1983zk}.

\section{One-dimension-less non-local effective action: a shortcut}

Let us now show we can get to the same effective action, and thence the same Casimir force per unit area, by relying on the gauge invariance. We will directly specify to the parallel plate geometry. The PMC conditions $n_{\mu}F_{\mu\nu}=0$ are gauge invariant themselves, so we can directly add them with the same set of (gauge invariant) multipliers as before, see \eqref{auxiliary_fields}, to the Maxwell action, yielding the gauge invariant action
\begin{eqnarray}\label{actionbis}
S_{E}&=&\int d^{4}x \left[\frac{1}{4}F_{\mu\nu}^2 + n_{\mu}F_{\mu\nu}B_{\nu} + n_{\mu}F_{\mu\nu}\bar{B}_{\nu}\right] \,.
\end{eqnarray}
Thanks to the shift symmetries $B_{\mu} \to B_{\mu} + \phi \, n_{\mu}$, (and similarly for $\bar B$), we can immediately choose $b_z=\bar b_z=0$. To integrate out the $A$-modes, we need to fix the gauge. As the action \eqref{actionbis} is manifestly gauge invariant, we can invoke the standard BRST quantization scheme of adding a BRST exact term to the action (or following the Faddeev-Popov ``trick'' as a special case thereof), without having to worry any further about boundary conditions or contributions. In the special case of the linear covariant gauge, we get
\begin{eqnarray*}
S_{E}&=&\int\!\!d^{4}x\! \left[\frac{1}{4}F_{\mu\nu}^2 + s\,s^{\dagger}(\frac{\xi}{2} c^\dagger c + \frac{1}{2} A_\mu A_\mu) +F_{zi}b_{i}+F_{zi}\bar{b}_{i}\right]
\end{eqnarray*}
which is invariant under the standard nilpotent BRST and anti-BRST transformations ($s^2={s^{\dag}}^{2}=0$) \cite{Ojima1980}
\begin{eqnarray*}
s A_{\mu} = -\partial_{\mu} c\,,\,\quad s c^{\dagger} = h\,&,&\,\quad s c = 0\,,\,\quad s h=0\;, \\
s^{\dagger} A_{\mu} = -\partial_{\mu} c^{\dagger}\,,\quad s^{\dagger} c^{\dagger} = 0\,&,&\,\quad s^{\dagger} c = -h\,,\,\quad s^{\dagger} h=0\;,
\end{eqnarray*}
where $h$ is the Nakanishi-Lautrup field. It is then most convenient to work in Feynman gauge $\xi=1$. After explicit integration over the $A$-modes, we arrive at the same $(2+1)$-dimensional gauge invariant effective model \eqref{action_effective}.

\section{The partition function and Casimir force}

Recognizing the transversal, respectively longitudinal, projectors
\begin{equation}\label{T_propagator}
T_{ij}(k)=\delta_{ij}-\frac{k_{i}k_{j}}{\mk^2}\,,\quad L_{ij}(k)=\frac{k_{i}k_{j}}{\mk^2}\;,
\end{equation}
the effective 3-dimensional theory written in momentum space has a local gauge redundancy under $b_{i}\to b_{i} + k_{i}\beta$, with $\beta$ an arbitrary function of $\vec{k}$, and similarly for $\bar b_i$. We will not bother here to derive the (non-local) effective action in configuration space, but we will fix the gauge in Fourier space by adding two terms of the form
$\tilde{b}_{i}\frac{k_{i}k_{j}}{|\vec{k}|^{2}}\tilde{b}_{j}$ and $\tilde{\bar{b}}_{i}\frac{k_{i}k_{j}}{|\vec{k}|^{2}}\tilde{\bar{b}}_{j}$,
\begin{eqnarray}\label{action_effective_eta}
 S[\widetilde{b},\widetilde{\bar{b}}] &=&  \frac{1}{2} \int\frac{d^{3}k}{(2\pi)^{3}} \frac{\mk}{8} \left[\widetilde{b}_{i}(k)T_{ij}(k)\widetilde{b}_{j}(-k)\right. \\
 &+&\left.\widetilde{b}_{i}(k)T_{ij}(k)\widetilde{\bar{b}}_{j}(-k)e^{-\mk L}+\widetilde{\bar{b}}_{i}(k)T_{ij}(k)\widetilde{b}_{j}(-k)e^{-\mk L}\right. \nonumber  \\
 &+& \widetilde{\bar{b}}_{i}(k)T_{ij}(k)\widetilde{\bar{b}}_{j}(-k)e^{-\mk L}+\eta\tilde{b}_{i}\frac{k_{i}k_{j}}{|\vec{k}|^{2}}\tilde{b}_{j} + \eta\tilde{\bar{b}}_{i}\frac{k_{i}k_{j}}{|\vec{k}|^{2}}\tilde{\bar{b}}_{j}  \;, \nonumber
\end{eqnarray}
where $\eta$ is a new gauge fixing parameter. Thence, compactly,
\begin{equation*}
 S[\widetilde{b},\widetilde{\bar{b}}] =  \frac{1}{2}\int\frac{d^{3}k}{(2\pi)^{3}} V_{i}^{T}(k)\,\mathds{D}_{ij}(k)\,V_{j}(-k) \;,
 \end{equation*}
with
\begin{equation*}
V_{i}(k) = \left(
        \begin{array}{c}
          \widetilde{b}_{i}(k) \\
          \widetilde{\bar{b}}_{i}(k) \\
        \end{array}
      \right) \;,\;
\mathds{D}_{ij}(k) = \frac{\mk}{8} \left(
                       \begin{array}{cc}
                         T_{ij} + \eta L_{ij}&  e^{-\mk L} T_{ij} \\
                      e^{-\mk L} T_{ij} &  T_{ij} + \eta L_{ij}  \\
                       \end{array}
                     \right) \;.
\end{equation*}

We can perform this Gaussian functional integral over the fields $\widetilde{b}$ and $\widetilde{\bar{b}}$,
\begin{equation*}
Z_{b}=\int \mathcal{D}\widetilde{b} \mathcal{D}\widetilde{\bar{b}} e^{S[\widetilde{b},\widetilde{\bar{b}}]}=\frac{C}{\sqrt{\det \mathds{D}}} \;,
\end{equation*}
where $C$ is an infinite constant not depending on $L$, by using the rule \cite{Peskin}
\begin{equation*}
(\det \mathds{D})^{-\frac{1}{2}} = e^{-\frac{1}{2} \int \frac{d^{3}k}{(2\pi)^{3}} \ln |\mathds{D}_{k}|} \;.
\end{equation*}

To compute the determinant of $\mathds{D}_{ij}(k)$, we observe that it is a $6\times6$ matrix for a specific vector value $\vec{k}$.  We find
\begin{equation*}
 |\mathds{D}_{k}| = \underbracket{\left(\frac{\mk}{8}\right)^{6} \eta^{2}} (1-e^{-2|\mk L})^2.
 \end{equation*}
The pieces above the bracket give a vanishing (and anyhow $L$-independent) contribution in dimensional regularization, so
\begin{equation*}
-\frac{1}{2} \int \frac{d^{3}k}{(2\pi)^{3}} \ln |\mathds{D}_{k}| = -\frac{1}{2} \int \frac{d^{3}k}{(2\pi)^{3}} \ln (1-e^{-2\mk L})^{2} = \frac{\pi^{2}}{720 L^{3}} \;.
\end{equation*}
The vacuum potential, defined via $Z_{b}=e^{-V_{L}}$, thus becomes $V_L=-\frac{\pi^{2}}{720 L^{3}}$ and the resulting finite force between the plates per unit area is (minus) the variation of $V_{L}$ with respect to $L$,
\begin{equation}\label{Casimir_force}
F=-\frac{\partial V_{L}}{\partial L}=-\frac{\pi^{2}}{240L^{4}}  \;.
\end{equation}
The result \eqref{Casimir_force} is the usual attractive Casimir force (see eq.~(9) of \cite{Bordag1998}), albeit here computed with PMC rather than with PEC (``perfect electric boundary conditions''). The equivalence between these two choices is however known in certain instances, as this case of infinite parallel plates, see \cite{Edery:2008rj}. Another computation of the Casimir force per unit area which might be simpler to extend to the non-Abelian case, also based on \eqref{action_effective}, is included in the Appendix \ref{extraapp}, making use of diagonalization.

The very same method can be applied to compute the Casimir force per unit area in $2+1$ dimensions with parallel lines separated by $L$. Indeed, going through the same steps, the $\mathds{D}$-matrix becomes then a $4\times4$ matrix and
\begin{equation*}
 |\mathds{D}_{k}| = \left(\frac{\mk}{8}\right)^{4} \eta^{2} (1-e^{-2|\mk L}).
 \end{equation*}
so that $V_L=-\frac{\zeta(3)}{16\pi L^{2}}$ and thus  $F=-\frac{\partial V_{L}}{\partial L}=-\frac{\zeta(3)}{8\pi L^{3}}$, the standard result \cite{Ambjorn:1981xw,Chernodub2018}. In fact, it is possible to obtain the formula for generic $d>1$ space-time dimensions\footnote{Note that, in principle, with this one dimension-less method, we need at least two space-time dimensions at the beginning for a meaningful result. For $d=2$, the Casimir force vanishes, as perhaps intuitively expected from the lack of propagating physical electromagnetic modes in that case. As a mathematical curiosity, the formula \eqref{Casimir_force_d} is finite and positive, viz.~equal to $\frac{1}{2L}$, for $d=1$.}. As
\begin{equation*}
|\mathds{D}_{k}| = \left(\frac{\mk}{8}\right)^{2(d-1)} \eta^{2} (1-e^{-2|\mk L})^{d-2}
\end{equation*}
leading to
\begin{eqnarray*}
V_{L} &=& \frac{1}{2} \int \frac{d^{d-1}k}{(2\pi)^{d-1}} (d-2)\ln (1-e^{-2\mk L})\nonumber\\ &=& - \frac{(d-2)}{2^{d}\pi^{\frac{d}{2}}L^{d-1}}\Gamma(\frac{d}{2}) \zeta(d) \;,
\end{eqnarray*}
and thence
\begin{equation}\label{Casimir_force_d}
F_{L} = -\frac{(d-2)(d-1)}{2^{d}\pi^{\frac{d}{2}}L^{d}}\Gamma(\frac{d}{2}) \zeta(d) \; \;.
\end{equation}
This result is also standard, and, in our conventions, the Casimir potential is $(d-2)$ times the Casimir potential for a massless scalar field \cite{Ambjorn:1981xw}.

\section{Periodic boundary conditions}

Let us now generalize our method to derive the Casimir force per unit area with the same geometrical configuration as before, but instead of imposing the conditions \eqref{Maxwell_vacuum_xi} to both plates, we now impose them just to one and identify
\begin{equation}\label{periodic_boundary_condition}
A_{\mu}\Big{|}_{z=L/2} = A_{\mu}\Big{|}_{z=-L/2} \;.
\end{equation}
These are {\it periodic boundary conditions} (PBC). This is equivalent to assuming there are no fields whatsoever outside the space-time between the plates ($z<-L/2$ and $z>L/2$). We thus consider only the space-time inside the plates, with action
\begin{equation*}
S_{E}=\int d^{3}x \, \int_{-L/2}^{L/2} dz \, \mathcal{L}_{E} \;.
\end{equation*}
This time we need just one set of fields $B^{i}=b^{i}(\vec{x})\delta(z-L/2)$ because PBC imply that, effectively, both plates are the same one\footnote{This statement can be seen also from the mathematical fact that, for PBC, $\delta(z-L/2)= \frac{1}{L}\sum\limits_{n} e^{-\frac{2\pi i n}{L}(z-L/2)} = \delta(z+L/2)$.}. However, we have to consider also the contribution from the gauge field $A_{\mu}$, and the ghost, anti-ghost fields $c^{\dagger}$, $c$, as now the integration ends depend on $L$.
After completing the squares to split the contributions of $b$ and $A$,
\begin{equation*}
Z=\int\mathcal{D}\widetilde{A'} \, e^{S[\widetilde{A'}]}\int\mathcal{D}\tilde{c}^{\dagger}\mathcal{D}\tilde{c} \, e^{S[\tilde{c}^{\dagger},\tilde{c}]}\int \mathcal{D}\widetilde{b} \, e^{S[\widetilde{b}]}=Z_{A}Z_{c}Z_{b} \;,
\end{equation*}
where now
\begin{eqnarray*}
k_{z} &\to& \frac{2\pi n}{L} \;,\quad n\in \mathds{Z} \;,\quad
\int \frac{dk_{z}}{2 \pi} \to\frac{1}{L} \sum\limits_{n} \;.
\end{eqnarray*}
Let us start with $S[\tilde{b}]$, which using the series results in the Appendix \ref{appendix_section_3}, can be written as,
\begin{equation}\label{S_b}
 S[\widetilde{b}] =  \frac{1}{2}\int\frac{d^{3}k}{(2\pi)^{3}} b_{i}^{T}(k)\,\mathds{D}_{ij}({k})\,b_{j}(-k) \;,
 \end{equation}
with
\begin{equation*}
\mathds{D}_{ij}({k}) = \frac{\mk}{8} \coth\left(\frac{L\mk}{2}\right)T_{ij}(k) \;.
\end{equation*}
As in the former case, the action \eqref{S_b} is a non-local action in momentum space, which has a gauge invariance. We can add a gauge fixing term $\propto\tilde{b}_{i}\frac{k_{i}k_{j}}{|\vec{k}|^{2}}\tilde{b}_{j}$ to make the eigenvalues of $\mathds{D}$ non-zero. With such a gauge fixing term,
\begin{equation*}
 |\mathds{D}_{k}| = \left(\frac{\mk}{8}\right)^{3} \eta\coth^2\left(\frac{L\mk}{2}\right)\;.
\end{equation*}
Again using dimensional regularization, we get
\begin{eqnarray*}V_L^{(b)}&=&
\frac{1}{2} \int \frac{d^{3}k}{(2\pi)^{3}} \ln |\mathds{D}_{k}|  = \frac{1}{(2\pi)^{3}} \int d^{3}k \ln~\coth\left(\frac{L\mk}{2}\right)\nonumber\\ &=&  \frac{1}{2\pi^{2}} \int_{0}^{+\infty} \, d\mk \, \mk^{2} \, \ln~\coth\left(\frac{L\mk}{2}\right) =\frac{\pi^{2}}{48L^{3}} \;.
\end{eqnarray*}
For computing the contributions of the gauge field $\widetilde{A}'$ and ghost, anti-ghost fields $c^{\dagger}$, $c$, we use $\zeta$-function regularization, concretely we rely on \cite{Canfora2015}, upon replacing the temperature $T=1/L$, and setting $r=0$. The $\widetilde{A}$-contribution becomes $ V_L^{(\widetilde A)}=-\frac{2\pi^2}{48L^3}$, while for the ghost and anti-ghost we get $V_L^{(c)}=\frac{\pi^2}{45L^3}$. The net Casimir force per unit area thus becomes $F=-\frac{\partial V_{L}}{\partial L}=-\frac{\pi^{2}}{240L^{4}}$, the same as in eq.~\eqref{Casimir_force}, although the space-time is different.

\section{Discussion}
The boundary effective action approach for the Casimir effect---summarized for the parallel plate case by \eqref{action_effective}---opens multiple portals to interesting generalizations. A first one will be the inclusion\footnote{Here $\tilde{F}_{\mu\nu}=\frac{1}{2}\epsilon_{\mu\nu\rho\sigma}F_{\rho\sigma}$ is the Hodge dual of the strength tensor $F_{\mu\nu}$ \cite{Nakahara}.} of the ``topological term'' $\propto \theta  \tilde F_{\mu\nu} F_{\mu\nu}$. Allowing $\theta$ to vary between the boundaries allows to model chiral media, see recent works like \cite{Jiang:2018ivv,Fukushima:2019sjn}, with unexpected sign flips in the Casimir energy/force. In e.g.~\cite{Fukushima:2019sjn}, gauge invariance is not manifest because the boundary conditions are put by hand on top of the equations of motion. Our approach is still based on a gauge invariant action and the action principle will naturally lead to a mixture of PMC and PEC, depending on the (variable) $\theta$. We will report on these results soon and compare with \cite{Fukushima:2019sjn}.

Another generalization will be to the non-Abelian case, where we foresee an interesting interplay between the Casimir energy and the non-perturbative effects generated by Gribov copies \cite{Gribov:1977wm,Capri:2015ixa}. The latter can be included in the path integration \emph{after} the inclusion of the boundary conditions as again, these are implemented in an explicitly gauge invariant fashion, on top of which a non-perturbative gauge (taking into account the copies) can be chosen. This could be of relevance for bag models' stability \cite{Chodos:1974je,Canfora:2013zna}.

\section*{Acknowledgments}
We thank F.~Canfora and I.F.~Justo for insightful discussions during the conception of this research. P.~P. is supported by Fondo Nacional de Desarrollo Cient\'{i}fico y Tecnol\'{o}gico--Chile (Fondecyt Grant No.~3200725), Charles University Project No.~UNCE/SCI/013, and was supported by the project High Field Initiative (CZ.02.1.01/0.0/0.0/15\_003/0000449) from the European Regional Development Fund during part of this work. D.~D and L.~R. are grateful for the hospitality at CECs (Valdivia, Chile) where this work was initiated and for the support from Fondecyt Grant No.~1200022.

\appendix

\section{Useful formulae}

\subsection{Fourier conventions}

\label{appendix_section_1}
We collect some useful expressions for the Fourier transformation of the gauge and auxiliary fields. Taking the convention of \cite{Peskin}, for the gauge field
\begin{equation}\label{A_Fourier}
A_{\mu}(x)=\int \frac{d^{4}k}{(2\pi)^{4}}\widetilde{A}_{\mu}(k)e^{-ik\cdot x} \;,
\end{equation}
where the integrals go from $-\infty$ to $+\infty$ for each momentum component, and $\widetilde{A}_{\mu}(k)$ is, by definition, the Fourier transformation of the gauge field $A_{\mu}(x)$, with dimension $L^{3}$ in natural units.

Considering that the fields $b_{i}(\vec{x})$ and $\bar{b}_{i}(\vec{x})$ are 3-vectors evaluated in 3-dimensional space,
\begin{equation}\label{b_Fourier}
b_{i}(x)=\int \frac{d^{3}k}{(2\pi)^{3}}\widetilde{b}_{i}(k)e^{-i\vec{k}\cdot\vec{x}} \;,
\end{equation}
$\widetilde{b}_{i}(k)$ is the Fourier transformation of the auxiliary field $b_{i}(x)$. We have an analogous expression for the auxiliary field $\bar{b}_{i}(x)$.

An expression for the Dirac delta-distribution in terms of its Fourier components is
\begin{equation}\label{delta_Fourier}
\delta(z\pm L/2)=\int \frac{dk}{2\pi} e^{-ik(z \pm L/2)} \;,
\end{equation}
where, once more, the integral over momentum $k$ goes from $-\infty$ to $+\infty$. For any function $f=f(z)$, we have
\begin{equation*}
\int dx  f(x) \delta(z \pm L/2) = f(\pm L/2) \;.
\end{equation*}
Note that, from a physical dimension point of view, $\delta(z \pm L/2)$ has dimension of momentum in natural units, as we can infer from \eqref{delta_Fourier}.

\subsection{Useful integrals}
\label{appendix_section_2}

In order to compute $S[\widetilde{b},\widetilde{\bar{b}}]$, some diverging  integrals must be treated. All of them can be obtained as a limiting case of the following formula:
\begin{eqnarray}\label{integral_generic}
I(L,y,s)&=&\int_{-\infty}^{\infty}{\frac{e^{-i x L}}{(x^2+y^2)^s} \frac{dx}{2\pi}}~~~\mbox{with}~~~L>0,~y>0,\\
&=&\frac{2^{\frac{1}{2}-s} \left(\frac{L}{y}\right)^{s-\frac{1}{2}}
   \mbox{K}_{s-\frac{1}{2}}(L y)}{\sqrt{\pi}\,\Gamma (s)},
\end{eqnarray}
which is convergent for $\text{Re}(s)>0$, using analytic properties of the Bessel function of the second kind $\mbox{K}_\alpha(x)$, and of the Gamma function $\Gamma(s)$. In a sense, it can be seen as a sort of zeta-regularization of the integrals.

In this manner we obtain, taking into account that $\mbox{K}_{1/2}(x)=\sqrt{\frac{\pi}{2x}}e^{-x}$, $\mbox{K}_{3/2}(x)=\sqrt{\frac{\pi}{2x}}e^{-x}(1+1/x)$ and $k^2={\mk^{2}+k_{z}^{2}}$,
\begin{eqnarray}\label{integral_kz_1}
\int \frac{dk_{z}}{2\pi} \frac{e^{-ik_{z}L}}{k^{2}}  &=&   I(L,\mk,1) 
= \frac{e^{-\mk L}}{2\mk} \;,  \nonumber\\
\int \frac{dk_{z}}{2\pi} \frac{k_{z}^{2} e^{-ik_{z}L}}{k^{2}}&=&-\frac{d^2 I(L,\mk,1)}{d L^2} 
=-\frac{1}{2} \mk e^{-\mk L} \;,   \nonumber\\
\int \frac{dk_{z}}{2\pi} \frac{k_{z}^{2} e^{-ik_{z}L}}{k^{4}}&=&-\frac{d^2 I(L,\mk,2)}{d L^2} 
=\frac{ e^{-\mk L}(1-\mk L)}{4\mk}\;,   \nonumber\\
\int \frac{dk_{z}}{2\pi} \frac{1}{k^{2}}&=&\lim_{ L\rightarrow0 } I(L,\mk,1)
= \frac{1}{2\mk} \;,\nonumber\\
\int \frac{dk_{z}}{2\pi} \frac{k_{z}^{2}}{k^{2}}&=&\lim_{
   L\rightarrow0 } -\frac{d^2 I(L,\mk,1)}{d L^2}
= -\frac{\mk}{2} \;, \nonumber\\
\int \frac{dk_{z}}{2\pi} \frac{k_{z}^{2}}{k^{4}}&=&\lim_{
   L\rightarrow0 } -\frac{d^2 I(L,\mk,2)}{d L^2}
= \frac{1}{4\mk} \;.
\end{eqnarray}
The interested reader may verify that in all cases, the same (finite) values are recovered in dimensional regularization.

\subsection{Useful series}
\label{appendix_section_3}

One of the series which appears when dealing with periodic boundary conditions is
\begin{equation}\label{series_convergent}
\sum\limits_{n=-\infty}^{+\infty}\, \frac{1}{\mk^{2}+(2\pi/L)^{2}n^{2}} = \frac{L \coth(\frac{\mk L}{2})}{2\mk} \;.
\end{equation}
The following series is not convergent, but it can be regularized:
\begin{eqnarray}\label{series_non_convergent}
&&\sum\limits_{n=-\infty}^{+\infty}\, \frac{(2\pi/L)^{2}n^{2}}{\mk^{2}+(2\pi/L)^{2}n^{2}}= \sum\limits_{n=-\infty}^{+\infty}\, \frac{-\mk^{2}+\mk^{2}+(2\pi/L)^{2}n^{2}}{\mk^{2}+(2\pi/L)^{2}n^{2}} \nonumber \\
 &&=  - \, \mk^{2}\frac{L \coth(\frac{\mk L}{2})}{2\mk} \, + \, \sum\limits_{n=-\infty}^{+\infty} 1   = -  \frac{L \mk \coth(\frac{\mk L}{2})}{2} \;, \nonumber
\end{eqnarray}
where the sum in the second equality has been dropped in dimensional regularization, since it is proportional to $\delta(0)=0$, see also footnote~5.

\section{Yet another confirmation}\label{extraapp}

To facilitate further computations (e.g.~when fermion loop corrections were to be added \cite{Bordag:1983zk}) and/or generalizations, it can be useful to decouple both sets of boundary modes appearing in the action \eqref{action_effective}. That is, we will diagonalize the relevant quadratic form by the following field transformation---with trivial Jacobian:
\begin{equation}\label{diag1}
  \widetilde b_i(k)= \widetilde f_i(k)\,,\qquad \widetilde{\bar b}_i(k)= -e^{-\mk L}\widetilde f_i(k)+ \widetilde{\bar f}_i(k)\,.
  \end{equation}
After a little algebra, the new effective action reads
\begin{eqnarray}
 S[\widetilde{f},\widetilde{\bar{f}}] &=&  \frac{1}{2} \int\frac{d^{3}k}{(2\pi)^{3}} \frac{\mk}{8} \left[\widetilde{f}_{i}(k)\left(1-e^{-2\mk L}\right)T_{ij}(k) \widetilde{f}_{i}(-k)\right.\nonumber\\&&\left.+\widetilde{\bar f}_{i}(k)T_{ij}(k) \widetilde{\bar f}_{i}(-k)\right]\,.
\end{eqnarray}
We choose dimensional regularization in $d=3-\epsilon$ dimensions. After proper gauge fixing and due to the absence of scale in the $\widetilde{\bar f}$-sector, the ensuing integration produces a null contribution to the vacuum energy. It remains to compute the $\ln\det$ coming from
\begin{eqnarray}
    &&\frac{1}{2} \int\frac{d^{3}k}{(2\pi)^{3}} \frac{\mk}{8} \left[\widetilde{f}_{i}(k)\left(1-e^{-2\mk L}\right)T_{ij}(k) \widetilde{f}_{i}(-k)\right.\nonumber\\&&\left.-\frac{1}{\alpha} \widetilde{f}_{i}(k) \mk \frac{k_i k_j}{\mk^2} \widetilde{f}_{i}(-k)\right]\,,
\end{eqnarray}
with $\alpha$ a gauge fixing parameter as before. The eigenvalues of the quadratic form are $(1-e^{-2\mk L}) \frac{\mk}{8}$ (multiplicity $d-1$), respectively $\frac{\mk}{\alpha}$ (multiplicity $1$), corresponding to the number of $d$-vectors transverse to $\vec k$, respectively parallel to $\vec k$. As such, we get by summing over $\ln(\text{eigenvalues})$
\begin{eqnarray}\label{diag3}
V_L&=& \frac{1}{2} \int \frac{d^d k}{(2\pi)^d}\left[ (d-1) \ln \left((1-e^{-2\mk L}) \frac{\mk}{8}\right)+ \ln  \frac{\mk}{\alpha}\right]
\nonumber\\ &=&\int \frac{d^3 k}{(2\pi)^3} \ln \left(1-e^{-2\mk L}\right)\quad \text{(dim.~reg.)}\nonumber\\&=&\frac{1}{2\pi^2}  \int_0^{+\infty} d\mk \mk^2\ln\left(1-e^{-2\mk L}\right)=-\frac{\pi^2}{720L^3}
\end{eqnarray}
which is the expected, gauge invariant, result.

\bibliographystyle{apsrev4-1}
\bibliography{Casimir_biblio}

\end{document}